\documentclass[pra,amsmath,amssymb,showpacs,superscriptaddress]{revtex4}
\usepackage{graphicx,epstopdf,bm}
\usepackage[normalem]{ulem}

\newcommand{\Fkt}[1]{\,\mathsf {#1}}
\ifx\Tr\renewcommand{\Tr}{\Fkt{Tr}}
\else\newcommand{\Tr}{\Fkt{Tr}}
\fi

\begin{document}

\title{Coherent control of bond making: The performance of
  rationally phase-shaped femtosecond laser pulses}
\thanks{dedicated to the memory of Moshe Shapiro}

\author{Liat Levin}
\affiliation{The Shirlee Jacobs Femtosecond Laser Research Laboratory,
  Schulich Faculty of Chemistry, Technion-Israel Institute of
  Technology, Haifa 32000, Israel}

\author{Wojciech Skomorowski}
\affiliation{Theoretische Physik, Universit\"at Kassel,
  Heinrich-Plett-Stra{\ss}e 40, 34132 Kassel, Germany
}

\author{Ronnie Kosloff}
\affiliation{Fritz Haber Research Centre and The
    Department of Physical Chemistry, Hebrew University, Jerusalem
    91904, Israel}

\author{Christiane P. Koch}
\affiliation{Theoretische Physik, Universit\"at Kassel,
  Heinrich-Plett-Stra{\ss}e 40, 34132 Kassel, Germany
}

\author{Zohar Amitay}
\affiliation{The Shirlee Jacobs Femtosecond Laser Research Laboratory,
  Schulich Faculty of Chemistry, Technion-Israel Institute of
  Technology, Haifa 32000, Israel}

\begin{abstract}
The first step in the coherent control of a photoinduced binary
reaction is bond making or photoassociation.
We have recently demonstrated coherent control of bond making
in multi-photon femtosecond photoassociation of hot magnesium
atoms, using linearly chirped pulses [Levin et al., arXiv:1411.1542].
The detected yield of photoassociated magnesium dimers was enhanced by
positively chirped pulses which is explained theoretically by
a combination of purification and chirp-dependent Raman transitions.
The yield could be further enhanced by pulse optimization resulting in
pulses with an effective linear chirp and a sub-pulse structure, where
the latter allows for exploiting vibrational coherences.
Here, we systematically explore the efficiency of phase-shaped pulses
for the coherent control of bond making, employing a parametrization
of the spectral phases in the form of cosine functions.
We find up to an order of magnitude enhancement of the yield compared
to the unshaped transform-limited pulse. The highly performing pulses
all display an overall temporally increasing instantaneous frequency
and are composed of several overlapping sub-pulses. The time delay
between the first two sub-pulses almost perfectly fits the vibrational
frequency of the generated intermediate wavepacket.
These findings are in agreement with chirp-dependent Raman transitions
and exploitation of vibrational dynamics as underlying control
mechanisms. 
\end{abstract}
\date{\today}
\pacs{42.65.Re, 82.50.Nd, 82.53.Eb, 82.53.Kp}
\maketitle

\section{Introduction}
Quantum coherent control of photo-induced
processes~\cite{TannorJCP86,Shapirobook,DantusCR04} 
directs a quantum system from its initial state to desired final
state(s) by manipulating interferences among coherent excitation
pathways. When using strong femtosecond laser pulses, control can be
realized---based on the broad spectrum, high intensity and
ultrashort duration of the pulses---by tailoring
their temporal shape~\cite{WeinerRSI00}. Coherent control using
tailored short pulses is nowadays successfully employed in many
different areas of quantum physics, including photoionization and
fragmentation,
nuclear magnetic resonance and quantum information
processing. However, the dream of coherently controlling
photo-induced bimolecular chemical
reactions~\cite{tannor1,TannorJCP86,KrauseJCP90,SeidemanCPL90}
which was at the origin of coherent control, still stands open.
For example, it is not yet understood under which conditions a
bimolecular reaction can be controlled and which reaction mechanisms
are particularly susceptible to control. This dream has attracted
significant interest over the years including that of Moshe
Shapiro~\cite{KrauseJCP90,SeidemanCPL90,ShapiroPRL96,AbrashkevichPRL98,ZemanPRL04}
since it holds the promise of creating a new type of photochemistry.

Two possible scenarios for the coherent control of a bimolecular
reaction have been discussed in the literature. The
first one relies on preparing coherent superposition states, in order
to determine the outcome of the
reaction~\cite{ShapiroPRL96,AbrashkevichPRL98,ZemanPRL04}.
Realizing this scenario requires control over reactive collisions
which currently seems to be extremely challenging. Given, however,
recent advances with cold~\cite{HensonSci12,HallPRL12}
and ultracold molecules~\cite{OspelkausSci10},
we may witness the advent of such a coherent chemistry in the not too
distant future.

The second scenario utilizes interference in time rather than
frequency domain, i.e., it is built on short, shaped pulses to induce
bond making (photoassociation) and breaking
(photodissociation)~\cite{RonnieDancing89}. The coherence of the
reactants required for controlling bond making must then be prepared
during the
photo-induced process. This is the reason why coherent control of a
binary reaction has proven so difficult. The quantumness of an
ensemble of reaction partners can be measured in terms of the purity,
$\mathcal P =\Tr[\rho^2]$, where $\rho$ describes the state of the reaction
partners, initially a thermal ensemble,
$\rho_0\sim\exp[-H/k_BT]$. Note that the purity is closely related to
the entropy of the sample; and thus the coherent interaction of the
reaction partners with the laser light does not change the purity.
Preparation of coherence can still be achieved by addressing a
subensemble of reactants. One possibility is given by filtering out
collision pairs in a small Franck-Condon window close to the repulsive
barrier of the potential, as demonstrated by us in multi-photon
femtosecond photoassociation of hot magnesium
atoms~\cite{RybakPRL11}. The selected collision pairs are correlated
due to the strong interaction close to the barrier. Correspondingly,
the subensemble of photoassociated, electronically excited molecules
shows coherent transient vibrational and rotational
dynamics that were identified using pump-probe femtosecond
spectroscopy~\cite{RybakPRL11}.
The generation of molecular coherence was shown to be possible,
even though the thermal ensemble of atoms was completely
incoherent, with an initial purity of essentially zero.
The generated coherences are amenable to coherent
control, as demonstrated by us very recently~\cite{LevinPRL15}: The yield
of detected magnesium dimers is enhanced by positively chirped pulses
and suppressed by negatively chirped pulses. Our {\it ab initio}
model has revealed that control is achieved by purification combined with
chirp-dependent Raman transitions.
Then, in closed-loop optimization experiments, an
improved pulse was obtained that displays an effective
temporal linear chirp and is
composed of two sub-pulses. The latter allows for utilizing  coherent
vibrational dynamics in addition to the chirp-dependent Raman
transitions~\cite{LevinPRL15}.

Here, we utilize the insight gained earlier~\cite{LevinPRL15} and
demonstrate an even higher degree of coherent control over bond making
in femtosecond photoassociation of hot magnesium atoms, achieving
up to an order-of-magnitude enhancement of the yield compared to the
unshaped transform-limited  pulse. The highly
performing phase-shaped pulses all display an overall temporally increasing
instantaneous frequency  and are composed of several overlapping
sub-pulses, where the time separation between the first two sub-pulses
almost perfectly fits the vibrational frequency of the generated
intermediate wavepacket. The corresponding spectral phases have the
general form of a cosine function. 
These findings can again be rationalized by chirp-dependent Raman
transitions and exploitation of vibrational dynamics as underlying
control mechanisms. 

\section{Excitation scheme and control signature}
\label{sec:scheme}
\begin{figure}[b]
  \centerline{\includegraphics[width=0.4\linewidth]{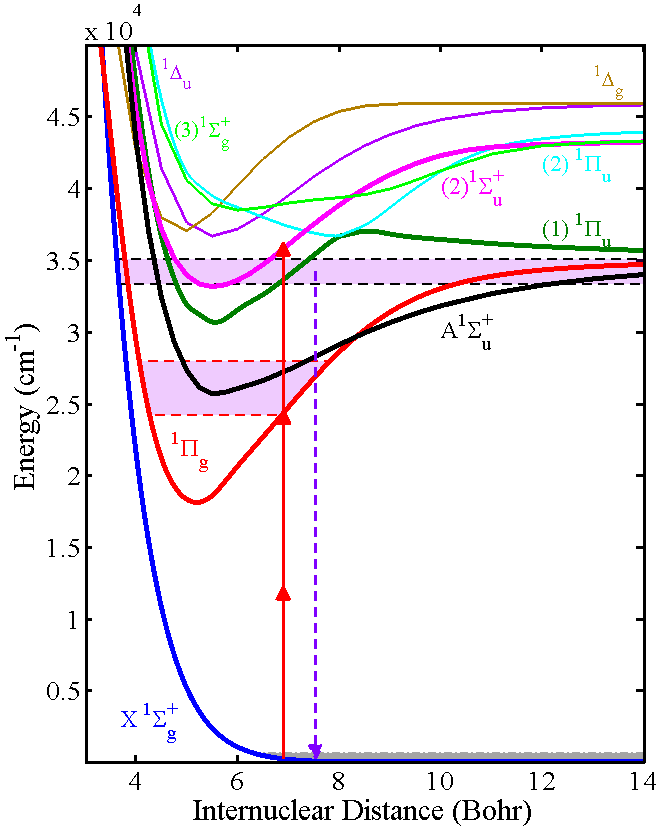}}
  \caption{The bond-making excitation scheme and potential energy
    curves of the magnesium     dimer.
    The excitation by three NIR photons is indicated by the red arrows
    and the UV emission by the dashed purple arrow.
    The shaded areas show the thermally populated $X$-state scattering
    states as well as
    the band of rovibrational levels after
    two-photon excitation and that from which UV emission is observed.
    }
\label{extplot}
\end{figure}
Pairs of magnesium atoms are photoexcited to form a magnesium dimer,
$\textrm{Mg}+\textrm{Mg} \rightarrow \textrm{Mg}^{\ast}_{2}$.
Figure~\ref{extplot} presents the bond-making excitation scheme,
starting from the van der Waals ground electronic state,
$X^1\Sigma^+_g$, and using a non-resonant two-photon transition to the
$(1)\Pi_g$ as a first step~\cite{RybakPRL11}.
Initially, the magnesium atoms, held at a temperature of 1000$\,$K,
thermally populate a manifold of $X$-state scattering states,
indicated by the blue-shaded area in Fig.~\ref{extplot}. The
excitation is driven by a strong, linearly polarized, shaped femtosecond pulse,
characterized by a transform-limited (TL) duration of 70$\,$fs and a
central wavelength of 840$\,$nm. The pulse energy corresponds to a TL
peak intensity of 7.2$\times$10$^{12}\,$W/cm$^{2}$.
Following the irradiation, atom pairs with short internuclear
separation undergo bond formation via a free-to-bound transition
into bound levels of a manifold of electronically excited states of
Mg$_{2}$. The gateway for this excitation is the excited state
$(1)^1\Pi_g$, and the corresponding range of rovibrational
levels that are populated in the $^1\Pi_g$ state is
indicated in Fig.~\ref{extplot} (in-between red dashed lines).

The photoassociated molecules are detected by measuring the intensity
of their UV fluorescence back to the ground electronic state, with a
lifetime of a few ns, in the wavelength range of 285-292$\,$nm.
This corresponds to UV emission from below the atomic asymptote
$^{1}P+^{1}S$ of excited Mg atoms (see the dashed blue arrow
schematically shown in Fig.~\ref{extplot}).
The corresponding range of UV-emitting levels which belong to the
$A^1\Sigma_u^+$, $(1)^1\Pi_u$, 
and  $(2)^1\Sigma_u^+$
electronically excited states, is also indicated in
Fig.~\ref{extplot} (in-between dashed black lines).

The total UV signal  originates from an incoherent
summation over the signals obtained for all thermally populated
initial scattering states which are characterized by their angular
momentum $J$ and collision energy $E_{coll}$.
We have previously shown~\cite{RybakPRL11,AmaranJCP13} that the main
contribution to the signal is from high partial waves, with $J$ around
70, where due to the centrifugal barrier the shallow van-der-Waals
potential well of the X-state supports only very few bound levels
or no bound levels at all.
This also implies that, as we have calculated~\cite{AmaranJCP13},
the bound-bound contribution to our signal corresponding to
excitation from bound levels in the $X$-state amounts to less than
20$\%$ of our total signal.

The bond-making excitation is controlled using
phase-shaped femtosecond pulses. Our earlier study~\cite{LevinPRL15}
employed  pulses with a linear chirp having
a spectral phase of the form
$\Phi(\omega)=\frac{1}{2} k (\omega - \omega_{0})^2$,
where $\omega_{0}$ is the central frequency and $k$ is the linear
chirp parameter. The best linearly chirped pulse, with
$k=0.006\,$ps$^2$, enhances the UV signal, compared to the TL pulse,
by a factor of 5.4.
Closed-loop optimization experiments have led to an improved pulse,
yielding an enhancement factor of 7.4, which features
an effective temporal linear chirp and is composed of two
sub-pulses.
Here, we investigate  two additional sets of shaped pulses,
all having a cosine spectral phase of the form
\begin{equation}
  \label{eq:cos}
  \Phi(\omega)=A \cos[(\omega-\omega_0) T + M]\,,
\end{equation}
characterized by  three parameters, $A$, $T$ and $M$.
Such a spectral phase temporally yields a sequence of sub-pulses
with a constant time difference $T$ between subsequent sub-pulses.
Once the value of $T$ is smaller than or comparable to the TL pulse
duration, 70$\,$fs in our case, the overall pulse results from several
overlapping sub-pulses that interfere with each other.
Hence, with $M$=0, and $A$ and $T$ having appropriate values,
the spectral phase of Eq.~\eqref{eq:cos} results in a
pseudo-chirped pulse. Such a pulse has a symmetric temporal intensity
profile with no splitting into separated sub-pulses. It therefore
can be assigned a well-defined pulse duration calculated as its
full width at half maximum. At the same time,
during the pulse's main intensity peak, the instantaneous
temporal frequency
increases, respectively decreases, monotonically.
Hence, a pseudo-chirped pulse can be characterized by pulse
duration and slope of the instantaneous frequency.
Examples of pseudo-chirped pulses are shown below in
Fig.~\ref{Ataumap}(a)-(f). For a given bandwidth,
a set of such pulses includes more combinations of pulse duration
and instantaneous frequency slope than a set of linearly chirped
pulses. This allows for a systematic investigation of the interplay of
chirp-dependent Raman transitions and exploitation of vibrational coherence
by temporally spaced pulses,
and is the first set of pulses we employ here.

The second set of shaped pulses we utilize
has cosine spectral phases with $M=5.7$.
With an additional proper choice for the values of $A$ and $T$,
the resulting shaped pulses are asymmetric in time
and effectively composed of several, partially overlapping
sub-pulses. Each sub-pulse has its own intensity and width, and
the time difference between subsequent sub-pulses can change across
the complete sequence.
These shaped pulses also exhibit an instantaneous frequency
that overall increases, respectively decreases, in time.
Examples of these asymmetric pulses
are found below in Fig.~\ref{AtaumapM57}(a)-(f).
For a given bandwidth, this set of pulses allows for a systematic
investigation of the role of the time delay between the sub-pulses as
well as their relative intensity.

\section{Experimental details and theoretical model}
\label{sec:details}
The magnesium photoassociation experiments are conducted in a heat pipe
with Ar buffer gas heated to 1000 K, in which the pressure of the
Mg vapor is around 5 Torr. Although a buffer gas is needed for the
stable and continuous operation of the heated cell over long time, it
does not play a part in the reaction. This is confirmed by observation
of a UV signal even without the buffer gas. Furthermore, the observed
UV signal remains the same when replacing argon with helium as a
buffer gas.
As we have previously discussed~\cite{LevinPRL15},
the UV signal can be ruled out to result from Mg ionization~\cite{GillenPRA01}.
%
%
Past experimental work shows that the existence of Mg clusters larger
than a dimer is highly negligible in our heat pipe~\cite{ZiemannZPD91}.
We therefore conclude that our measured signal results only from
excitation of magnesium dimers.

The shaped femtosecond laser pulses irradiate the sample at a
repetition rate of 1$\,$kHz, after undergoing shaping using
a liquid-crystal spatial light phase modulator~\cite{WeinerRevSciInst00}.
The measured UV fluorescence signal is collected at the
laser beam entrance to the heat pipe, at a small angle from the beam axis,
using a dichroic mirror, filter and lens, focusing the UV light onto
an optical fiber that is coupled to a spectrometer
and a time-gated camera system with a 5$\,$ns gate.

We employ the theoretical framework  developed in
Refs.~\cite{RybakPRL11,AmaranJCP13,RybakFaraday11,LevinPRL15} to describe
photoassociation of hot atoms by an intense laser field.
Our first principles  model combines {\it ab initio} electronic structure
theory for the Mg$_2$ molecule with quantum molecular dynamics to
describe the light-matter interaction in a non-perturbative
way. Random phase wavefunctions are employed to account for thermal
averaging~\cite{AmaranJCP13}.
In brief, we  explicitly include all electronic states shown in
Fig.~\ref{extplot}. The $X^1\Sigma_g^+$ ground electronic state is
coupled to the $(1)^1\Pi_g$ state by a two-photon coupling. Then, the
$(1)^1\Pi_g$ state is coupled by resonant one-photon couplings with a several
states of {\it ungerade} symmetry, the most important of which are
$A^1\Sigma_u^+$, $(1)^1\Pi_u$ and $(2)^1\Sigma_u^+$.
In addition, there are near-resonant one-photon couplings with the
$^1\Delta_{u/g}$ states. Moreover, since the system is exposed to a strong
field, we also include the Stark shifts due to one-photon couplings
with off-resonant states.
The strong nonadiabatic interaction between the first two
$^1\Pi_u$ states is taken into account by transformation  into the
diabatic basis.  The general structure of the
Hamiltonian is given in Eq.~(12) of
Ref.~\cite{AmaranJCP13}; it is augmented here by the additional states
shown in Fig.~\ref{extplot} and their one-photon couplings to the
$(1)^1\Pi_g$ state. The time-dependent Schr\"{o}dinger equation is solved
using the Chebyshev propagator~\cite{chebyprop} separately for
each thermal random-phase  basis  wavefunction  and rotational quantum
number $J$~\cite{AmaranJCP13}.
Any coupling between rotational and vibrational degrees of freedom is
neglected to ease the computational effort. Time-dependent, thermally
averaged
expectation values are calculated by proper weighting for each partial
wave, cf. Eqs.~(25) and~(26) of Ref.~\cite{AmaranJCP13}.
When the propagation is over, the final populations of the bound
levels in the $A^1\Sigma_u^+$, $(1)^1\Pi_u$ and $(2)^1\Sigma_u^+$
excited states are used to generate the integrated emission spectrum,
which is the experimental observable. For each electronic state,
the emission spectrum is obtained
by means of the Einstein coefficients with appropriate transition
dipole moments and including the line broadening.

\section{Results and discussion}
\label{sec:results}

\subsection{Pseudo-chirped pulses}
\label{subsec:pseudochirped}
\begin{figure}[b] 
  \centerline{\includegraphics[width=0.8\linewidth]{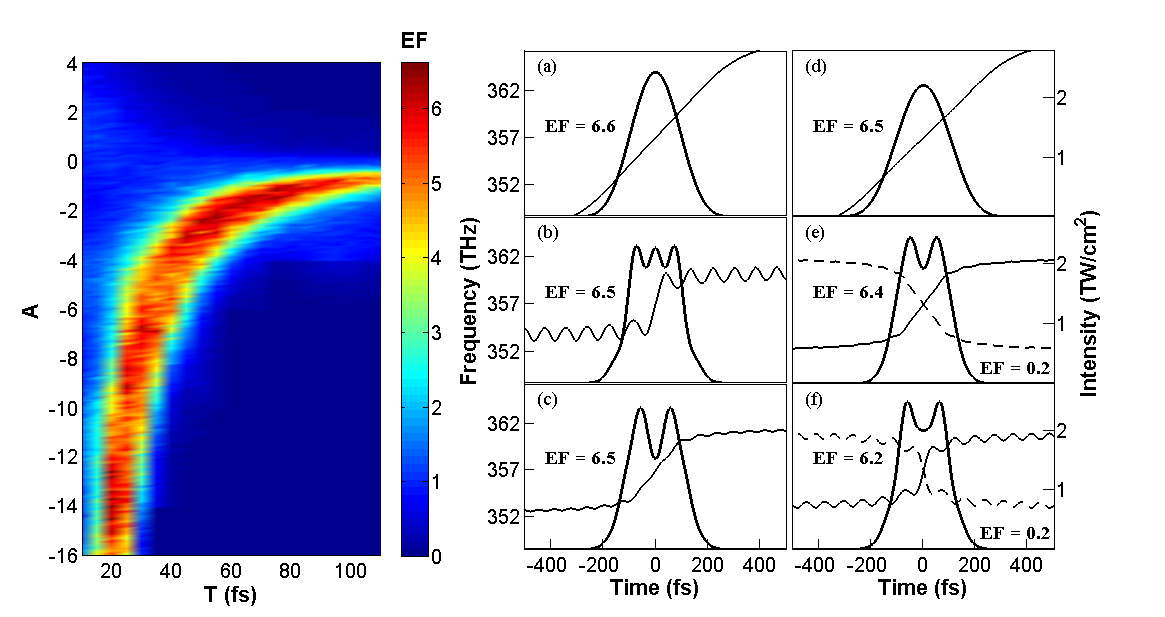}}
  \caption{Pseudo-chirped pulses ($M=0$).
    Left: Experimental UV signal enhancement relative to the TL pulse (EF) as a
    function of $T$ and $A$, cf. Eq.~\eqref{eq:cos}.
    Right: Examples of pseudo-chirped
    pulses yielding large enhancement. The intensity profile in time
    domain and the corresponding instantaneous frequency are plotted for:
    (a) $A=-12.6$, $T=20$, 
    (b) $A=-1.5$,  $T=75$, 
    (c) $A=-2.3$,  $T=55$, 
    (d) $A=-14$,   $T=20$, 
    (e) solid line:  $A=-2.4$, $T=50$, 
        dashed line: $A=2.4$,  $T=50$,
    (f) solid line:  $A=-1.7$, $T=65$, 
        dashed line: $A=1.7$,  $T=65$.
    The corresponding values of the measured enhancement (EF) are also indicated.}
  \label{Ataumap}
\end{figure}
We first discuss the case $M=0$ in Eq.~\eqref{eq:cos}.
Figure~\ref{Ataumap} presents the measured enhancement factor (EF), i.e., the
ratio of the UV signal for the shaped pulse over the signal for the
TL pulse, as a function of the phase shape parameters $T$ and $A$.
The maximal enhancement for cosine phase shapes with $M=0$ is found
to be $6.6$, slightly larger than the value for linearly chirped
pulses. For positive values of $A$, we find suppression of the signal
(EF$<$1), similarly to the results for linearly chirped pulses with
negative chirp~\cite{LevinPRL15}. The right-hand side of
Fig.~\ref{Ataumap} displays examples of pseudo-chirped pulses which
produce a large enhancement.

\begin{figure}[tbp]
  \centerline{\includegraphics[width=0.5\linewidth]{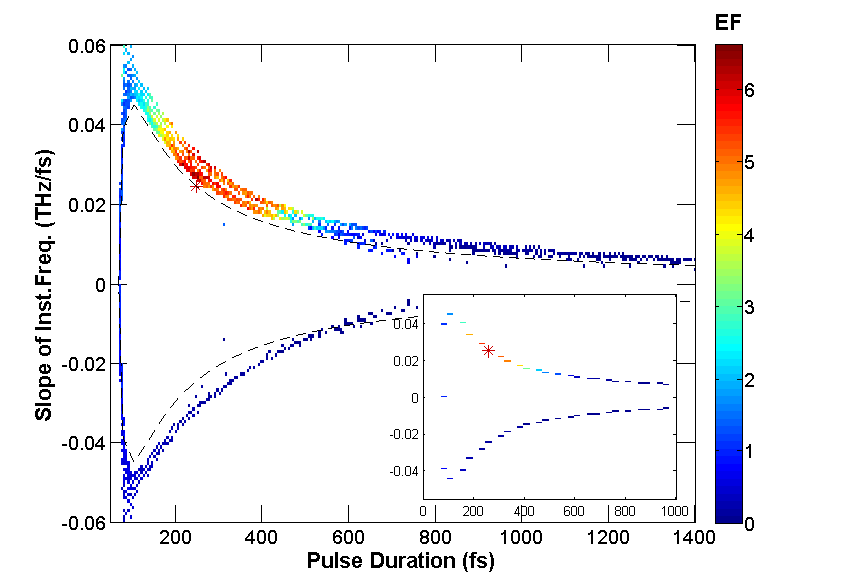}}
  \caption{Experimental UV signal enhancement (EF) as a function of instantaneous
    frequency slope and pulse duration (over the central part of the
    pulse) obtained for pseudo-chirped pulses.
    Inset: The corresponding map obtained for linearly chirped pulses.
    The dashed lines in the main map present the duration-slope combinations
    of the linearly chirped pulses (plotted to guide the eye).
    }\label{slopesdursmap}
\end{figure}
As explained above in Section~\ref{sec:scheme}, pseudo-chirped pulses
can be characterized by their duration and slope of the instantaneous
frequency. Translating the spectral phase parameters $A$
and $T$ into these characteristics,
Fig.~\ref{slopesdursmap} shows the enhancement factor as a function
of pulse duration and temporal frequency slope.
For comparison the inset shows the same map for linearly chirped
pulses. Clearly,
the range of pulse durations between approximately $200\,$fs and 400$\,$fs
and instantaneous frequency slopes between approximately
$0.01\,$THz/fs and $0.04\,$THz/fs
is optimal for achieving large
enhancement factors, between 4.5 to 6.6. This finding is similar
for both cosine phase-shaped pulses and linearly chirped pulses.
For the highest enhancement factors, above 6.0,
the optimal range of pulse durations and frequency slopes becomes narrower,
with the corresponding values of
200$-$280$\,$fs and 0.02$-$0.03$\,$THz/fs, respectively.
The dependence on the sign of the chirp, respectively slope of the
instantaneous frequency, is reflected in the asymmetry of the map with
respect to the axis of zero slope: For negative slope, suppression
rather than enhancement (EF$<1$)
is found. The corresponding pulses
are shown by dashed lines in Fig.~\ref{Ataumap}(e),(f).

The results show that there is an extended set of chirped and pseudo-chirped pulses,
with an extended range of pulse durations and frequency slopes,
that lead to a large enhancement of the detected photoassoication yield.
We associate the significant tolerance to the exact shape of the optimal pulse
with the fact that, due to the initial thermal conditions,
the photoassociation takes place from
many different initial scattering states of the colliding atoms.
Since each such initial state generally has a different energy and a slightly different wavefunction,
the corresponding optimal pulse has a slightly different shape.

\subsection{Asymmetric pulses with a sub-pulse structure}
\label{subsec:asymmetric}
\begin{figure}[tbp]
  \centerline{\includegraphics[width=0.8\linewidth]{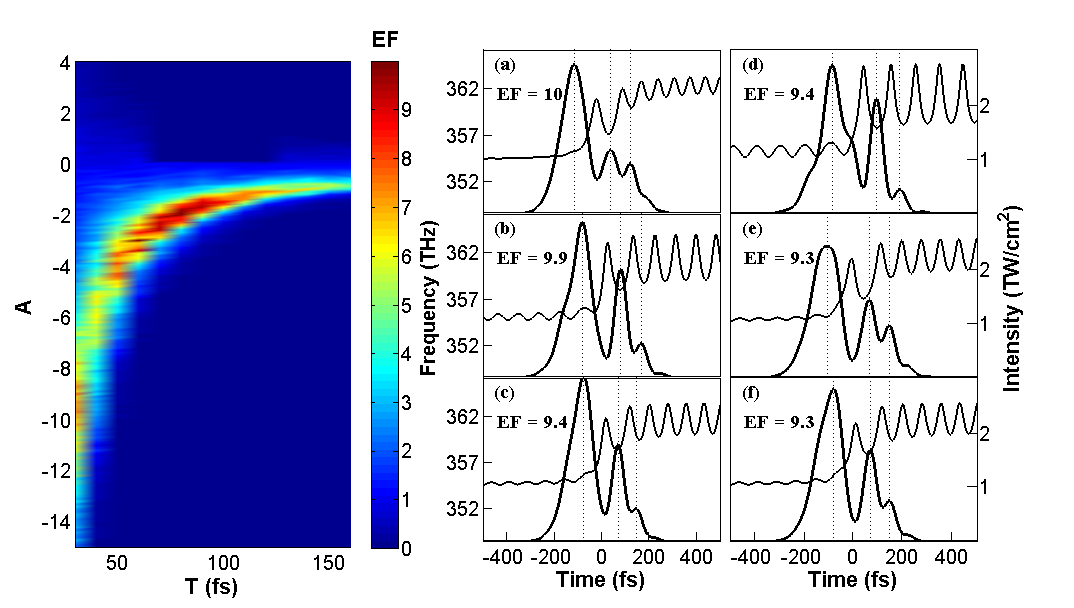}}
  \caption{Asymmetric pulses with a sub-pulse structure ($M=5.7$).
    Left: UV signal enhancement (EF) as a function of $T$ and
    $A$, cf. Eq.~\eqref{eq:cos}. Right: Examples of asymmetric pulses with a sub-pulse structure
    yielding large enhancement. The intensity profile in
    time domain and the corresponding instantaneous frequency are plotted
    for: (a) $A=-3$, $T=60$, (b) $A=-2$, $T=80$, (c) $A=-2.1$, $T=70$,
    (d) $A=-1.7$, $T=90$, (e) $A=-2.6$, $T=70$, (f) $A=-2.3$, $T=70$.
    The corresponding values of the measured enhancement (EF) are also indicated.}
  \label{AtaumapM57}
\end{figure}
The second set of cosine phase-shaped pulses uses $M=5.7$ in
Eq.~\eqref{eq:cos}, leading to pulses with a sub-pulse structure.
The experimental enhancement as a function of
the parameters $T$ and $A$ is presented in the left-hand side of
Fig.~\ref{AtaumapM57}, together with six pulses producing a
large enhancement over TL pulses in the right-hand side.
The best asymmetric pulses result in higher enhancement factors than
the best pseudo-chirped pulses. Here, the enhancement reaches an order
of magnitude compared to TL pulses and is larger
by about a factor of $1.5$ compared to the pseudo-chirped pulses.

\begin{figure}[tbp]
  \centerline{\includegraphics[width=0.6\linewidth]{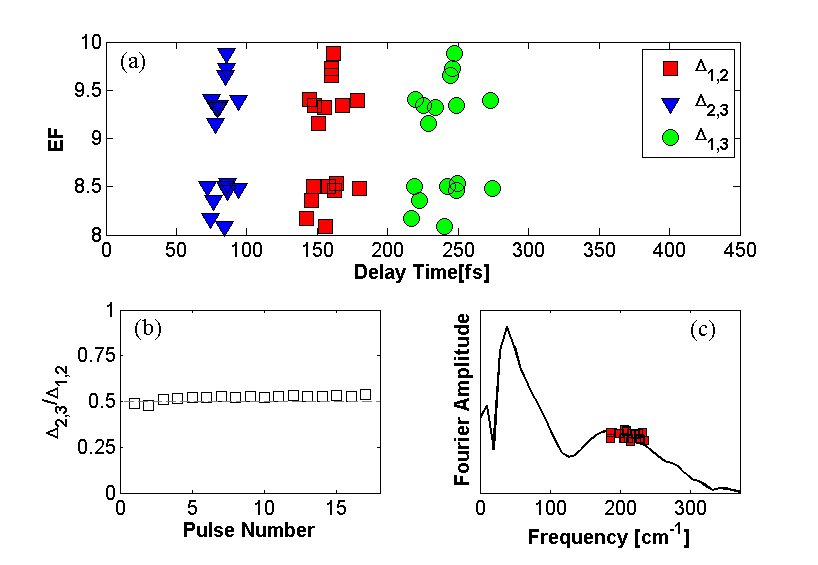}}
  \caption{(a) Enhancement factor vs time delays between sub-pulses:
    $\Delta_{1,2}$, i.e., delay between the first and second
    sub-pulse (red), $\Delta_{2,3}$ (blue), and $\Delta_{1,3}$
    (green).
    (b) Time delay ratio $\Delta_{1,2}/\Delta_{2,3}$ for the data
    shown in panel (a). The average value corresponds to 0.52.
    (c) Spectrum of the quantum beat frequencies obtained in the
    pump-probe experiment~\cite{RybakPRL11} and
    frequencies corresponding to the delays shown in panel (a),
    which fit the vibrational peak of the quantum beat spectrum.}
\label{EF_above8onQB}
\end{figure}
The duration of these pulses is more difficult to determine since they
are divided into several sub-pulses. A better understanding of the
advantage of these pulses over pseudo-chirped and linearly chirped
pulses is obtained by inspecting the time delay between the sub-pulses.
The optimal time delays between sub-pulses are identified
by considering only those asymmetric pulses that produce
enhancement factors larger than $8$, see Fig.~\ref{EF_above8onQB}(a).
For the delay between the first two sub-pulses ($\Delta_{12}$), the optimum is found
to be around $160\,$fs (red squares in
Fig.~\ref{EF_above8onQB}(a)).
The optimal delay between the second and third
sub-pulse ($\Delta_{23}$) is around $80\,$fs (blue squares in
Fig.~\ref{EF_above8onQB}(a)), consistent with half of the
first delay, see also Fig.~\ref{EF_above8onQB}(b).
The delay between the first two sub-pulses corresponds to a
frequency of $\sim210\,$cm$^{-1}$. Using femtosecond pump-probe
spectroscopy, we have earlier determined the quantum beat frequencies
relevant in the photoassociation process~\cite{RybakPRL11}.
The corresponding Fourier spectrum of the experimental transient pump-probe signal
is compared in
Figure~\ref{EF_above8onQB}(c) to the frequencies corresponding to the
time delays of the asymmetric pulses. The delay between the first two
sub-pulses (red squares) is found to fit the vibrational part of the
pump-probe signal.
The delay $\Delta_{13}$ between the first and third sub-pulse provides
a rough estimate of the overall pulse pulse duration. For the well
performing asymmetric pulses of Fig.~\ref{EF_above8onQB}(a),
$\Delta_{13}$ is found to be in the  range of 220$\,$fs to 275$\,$fs. This is
in good agreement with the optimal pulse durations of
the well performing pseudo-chirped pulses reported
in Sec.~\ref{subsec:pseudochirped}.

\begin{figure}[tb]
  \centerline{\includegraphics[width=0.55\linewidth]{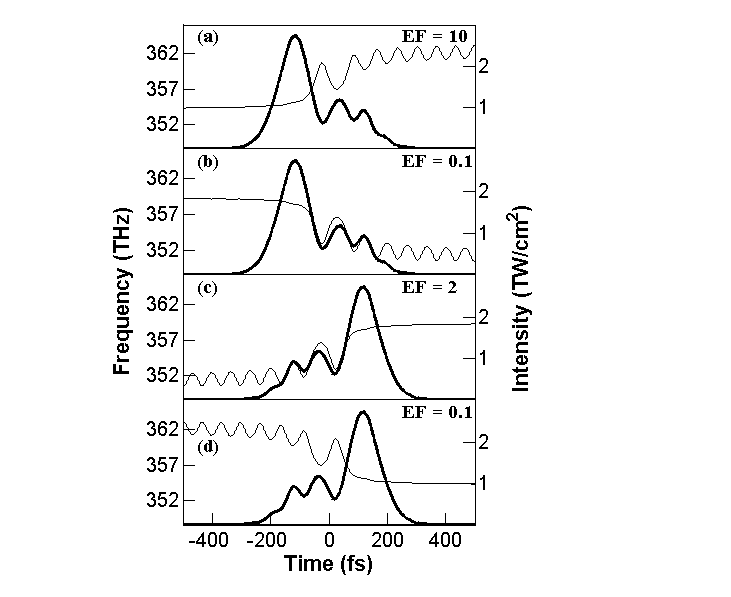}}
  \caption{Comparison of asymmetric pulses with a sub-pulse structure
    that have related pulse shape,
    obtained by different phase shape parameters ($A$, $T$, $M = \pm 5.7$),
    and their experimental EF values.
    Panels (a) and (b), respectively
    (c) and (d) compare pulses with the same temporal intensity profile but
    overall increasing vs. overall decreasing instantaneous
    frequency.
    Additionally, the relative intensity between the
    sub-pulses is compared
    between (a), (b) on one hand and (c), (d) on the other.
    (a) $A=-3$, $T=60$, $M=5.7$,  (b) $A=3$, $T=-60$, $M=5.7$,
    (c) $A=-3$, $T=60$, $M=-5.7$, (d) $A=3$, $T=60$, $M=5.7$.
  }  \label{pulsesfreqsubscomp}
\end{figure}
All well performing asymmetric pulses, cf. Fig.~\ref{EF_above8onQB}(a)
and examples in Fig.~\ref{AtaumapM57}, have a first sub-pulse that is
of higher intensity and longer duration than the subsequent
sub-pulses. These pulses also display  an overall increasing instantaneous
frequency  where the onset of the increase is located in the second
part of the first sub-pulse.
The role of delay and instantaneous frequency is further analyzed in
Fig.~\ref{pulsesfreqsubscomp} which displays four pulses together
with the enhancement that they yield. Comparison of panels (a) and
(b), respectively (c) and (d), which show pulses with the same
temporal intensity profile but overall increasing vs. overall decreasing
instantaneous frequency, confirm the earlier observation of a
clear advantage of increasing compared to decreasing instantaneous
frequency: An overall increasing (decreasing) instantaneous frequency
results in  enhancement (suppression) of the UV signal.
Figure~\ref{pulsesfreqsubscomp} also allows to examine the role of
relative intensity between the sub-pulses. Comparing, for example,
panels (a) and (c) which both feature an overall increasing
instantaneous frequency but differ in the increase vs decrease of the
sub-pulse peak amplitudes, reveal the advantage of a strong first
sub-pulse: The enhancement in panel (a) is larger by a factor of 5
compared to~(c).

Let us summarize the features of asymmetric pulses that are
particularly advantageous for enhancing the observed UV signal:
(i) a stronger and longer first sub-pulse,
(ii) an overall increasing instantaneous frequency which starts to
increase in the second part of the first sub-pulse,
(iii) an optimal sub-pulse delay $\Delta_{12}$ fitting exactly the $(1)^1\Pi_g$
vibrational period, and
(iv) an optimal sub-pulse delay $\Delta_{23}$ fitting half the $(1)^1\Pi_g$
vibrational period.
These findings suggest the following interpretation:
The first sub-pulse creates a $(1)^1\Pi_g$ vibrational wavepacket,
most likely at the right turning point,
via free-to-bound two-photon photoassociation transitions.
The non-resonant nature of these transitions requires a strong and
temporally wide sub-pulse.
In the second part of the first sub-pulse, the frequency starts to
increase  such that, after half a vibrational period,
when the wavepacket has reached the left turning point,
part of the amplitude is efficiently transferred into lower vibrational
levels of the $(1)^1\Pi_g$ state by resonant Raman transitions via
higher electronically excited states.
These Raman transitions are favored by an instantaneous frequency
increase~\cite{LevinPRL15}.
Then, after another half vibrational period,
i.e., one vibrational period after the photoassociation transition,
the wavepacket has reached again the right turning point. This
coincides with the peak amplitude of the second sub-pulse, i.e., with
$\Delta_{12}$, such that
another part of the wavepacket, larger than before,  is efficiently
and coherently excited into lower $(1)^1\Pi_g$ vibrational levels via
up-chirped Raman transitions.
Last, after one more half vibrational period ($\Delta_{23}$),
the wavepacket is again at the left turning point, where the third
sub-pulse drives another coherent Raman transition into lower
$(1)^1\Pi_g$ vibrational levels.
As we have previously shown and discussed~\cite{LevinPRL15},
the coherent enhancement of population in the
lower vibrational levels of the $(1)^1\Pi_g$ state is directly
reflected in the final population of the observed UV emitting band,
which is accessed from the former by a weak one-photon
transition.

We now compare these experimental findings to simulation results.
For the two most efficient pulses, shown in Fig.~\ref{AtaumapM57}(a)
and~(b), with experimental EF of 10 and 9.9, the theoretically
obtained enhancement amounts to 6.8 and 18.3,
respectively. Qualitatively, this confirms the strong enhancement of
the UV signal for phase-shaped pulses compared to the TL pulse. In
particular, the largest theoretically found enhancement for cosine
phase-shaped pulses is also larger than the maximum enhancement
obtained with linearly chirped pulses.
However, due to the sensitivity of the dynamics on the details of
the electronic structure data, as discussed in
Ref.~\cite{LevinPRL15}, it would be very challenging to reproduce the
exact values for the experimental enhancement
factors. In particular, the maximum enhancement was found for
larger chirp rates in the calculations than in
experiment~\cite{LevinPRL15}. Moreover, 
the agreement between experimental and
theoretical results for the vibrational spectra
of the pump-probe transients was satisfactory but not
perfect~\cite{RybakPRL11}. 
A pulse which is experimentally found to
maximally exploit the vibrational dynamics will therefore not
necessarily be the best performing pulse in the calculations where the
vibrational dynamics is slightly different.

\begin{figure}[tb]
  \includegraphics[width=0.6\linewidth]{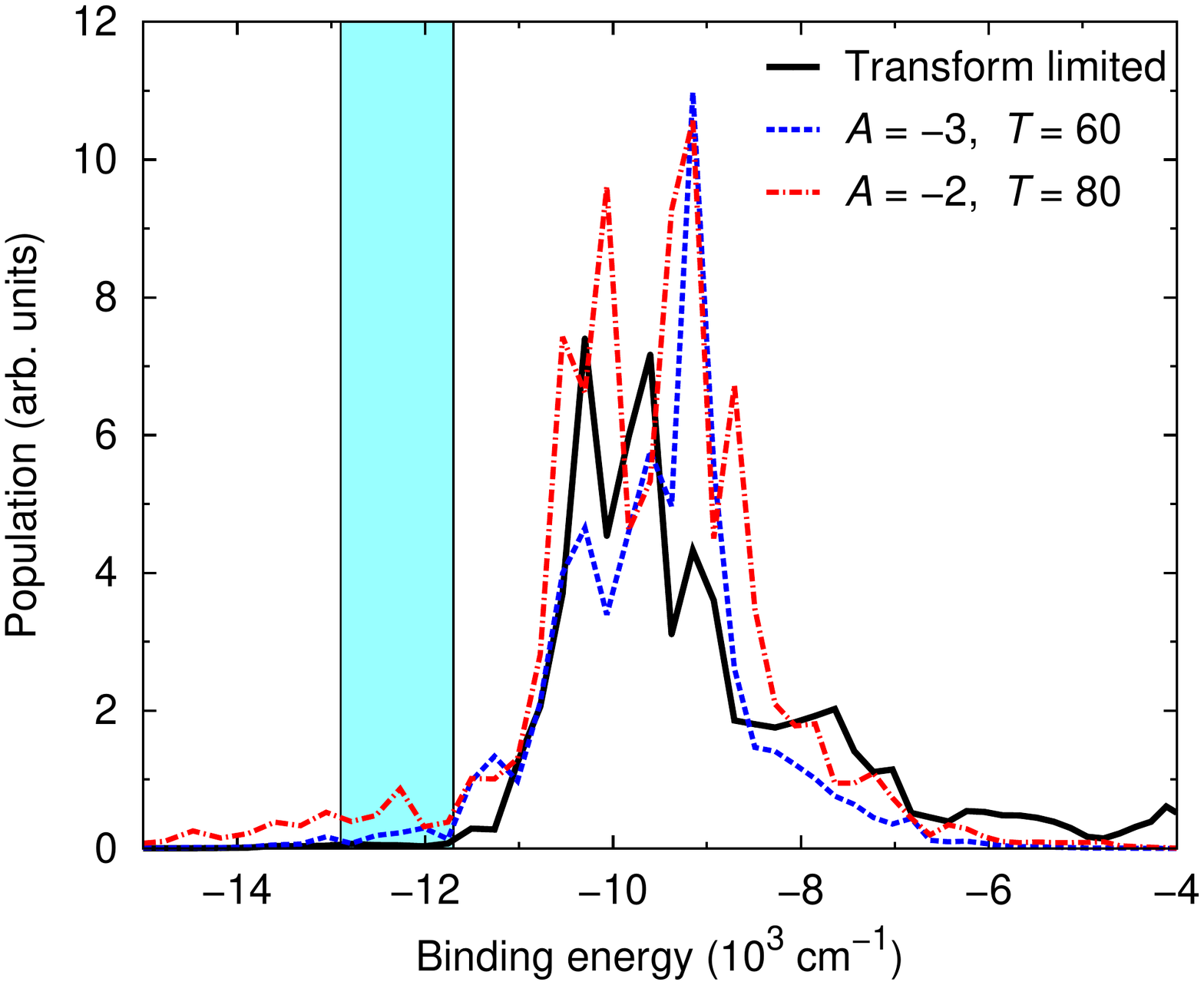}
  \caption{Final distribution of vibrational levels in
    intermediate $(1)^1\Pi_g$ state for two optimal pulses compared
    with a TL pulse. The shaded area shows the band which can be
    transfered by a one-photon transition to the UV emission window
    below the $^1S+ {^1P}$ atomic threshold.}
  \label{Pigvib}
\end{figure}
Figure~\ref{Pigvib} presents the final vibrational distribution in
the intermediate $(1)^1\Pi_g$ state. Assuming
that the final probe transition from the $(1)^1\Pi_g$
state into the UV emitting states can be treated
perturbatively, it is the
$(1)^1\Pi_g$ vibrational distribution that reveals the control
mechanism~\cite{LevinPRL15}. For example, linearly chirped pulses were
found to up-, respectively down-shift the $(1)^1\Pi_g$ vibrational
distribution by Raman transitions between  the $(1)^1\Pi_g$ and higher
lying electronic states.
This modifies in particular the population that is
transferred by the probe photon, marked by the shaded area in
Fig.~\ref{Pigvib},  into the band of rovibrational levels
in the \textit{ungerade}-states that produce the measured UV
emission~\cite{LevinPRL15}.
Inspection of Fig.~\ref{Pigvib} clearly shows that
cosine phase-shaped pulses tend to transfer more population to the
most deeply bound levels in the $(1)^1\Pi_g$ band than the TL pulse.
Thus, the Raman cycling process discussed in Ref.~\cite{LevinPRL15} is
efficient also in case of the new family of asymmetric
sub-pulse structured  pulses.

\section{Conclusions}
\label{sec:concl}

Control of binary reactions is one of the most difficult tasks of
coherent control. The first step of two-photon Franck Condon filtering
purifies the initial thermal state and thus sets the stage for
control. The target task in the present study was to control the
generation of bound Mg$_{2}$ molecules in the band of vibrational
levels of \textit{ungerade} states just below the $^1$P+$^1$S
dissociation threshold.
Population in this band is detected experimentally by its transition
back to the ground state which results in emission of UV light.
Once a target is set we can decipher the interference control
mechanism optimizing this task. The present work demonstrates the
power of a constructive approach based on cosine spectral phase
patterns. The constructive approach limits the number of control
parameters and therefore enables a rational approach to optimization.
We are able to identify two control mechanisms---electronic and
vibrational. The electronic component employs the full manifold
of electronic surfaces generating interference pathways which are
Raman-like. The electronic control mechanism enhances population in
intermediate levels which are transfered to the target band by a weak
final probe step. In addition, we identify by the time delay between
sub-pulses a vibrational mechanism which employs the interference
between a direct excitation and a delayed excitation.
We find a relatively high tolerance to the exact shape of the optimal
pulse for enhancing the detected UV signal.
We associate it with the highly thermal character of the initial
atomic ensemble which is held at a temperature of 1000$\,$K.

The roadmap explored in the control of strong-field photoassociation
of magnesium dimers opens the door to more complex binary
reactions. The main lesson learnt with Mg$_2$ can be summarized as
follows: Bond making requires a purification step which implies that
only a relatively small number of reactants can be addressed.
It is then crucial 
to employ shaped femtosecond pulses of high intensity
for maximizing this number as much as possible and, at the same time,
coherently controlling the excitation dynamics to a very high degree.
%
On the other hand, it is also crucial
to ensure an almost background-free detection of the photoassociated molecules. 
It was achieved here by the choice of closed-shell reactants and
a non-resonant multiphoton photoassociation transition.
This insight provides a blueprint for studying e.g. triatomic reactions which
would allow for stable reaction products and two possible reaction
outcomes. The basic scenario of utilizing interference in time domain
for the coherent control of binary reactions has thus been
established.
%
%
Thirty years after conception of the idea, perspectives for
the coherent control of binary reactions finally look bright.

\begin{acknowledgments}
This research was supported by The Israel Science Foundation (Grant
No.~1450/10) and the Alexander von Humboldt Foundation (WS).
\end{acknowledgments}


\end{document}